\documentclass[journal=nalefd,manuscript=article]{achemso}
\usepackage{graphicx}			
\usepackage{bm}				
\usepackage{amsmath}			
\usepackage{multirow}
\newcommand*\degree{$^{\circ}$}

\author{Priyamvada Jadaun}
\email{priyamvada@utexas.edu}
\author{Hema C. P. Movva}
\author{Leonard F. Register}
\author{Sanjay K. Banerjee}
\affiliation[UT Austin]
{Microelectronics Research Center, The University of Texas at Austin, Austin, TX 78758}

\title[ICl-GIC, IBr-GIC]
{Theory and synthesis of bilayer graphene intercalated with ICl and IBr for low power device applications}

\abbreviations{DFT, ICl, IBr, FET}
\keywords{graphene transistors; intercalation; density functional theory}

\begin{document}

\begin{abstract}
Graphene intercalation materials are potentially promising for the implementation of the ultra-low power, excitonic-condensate-based Bilayer pseudoSpin Field-Effect Transistor (BiSFETs) concept, as well as other novel device concepts requiring a graphene interlayer dielectric. Using density functional theory (DFT) we study the structural and electronic properties of bilayer graphene intercalated with iodine monochloride (ICl) and iodine monobromide (IBr). We determine the structural configuration of ICl and IBr graphene intercalation compounds (GICs). We also conduct an in-depth exploration of inter-layer electronic coupling, using \textit{ab initio} calculations. The presence of intercalants dopes the graphene layer. It also reduces, but does not eliminate, the electronic coupling between graphene layers, which may enable BiSFET operation. In addition, we present experimental results for ICl-GIC synthesis and characterization. 
\end{abstract}

\section{Introduction}

	Graphene has generated keen interest\cite{geim_rise_2007} for novel electronic applications such as the Bilayer pseudoSpin Field-Effect Transistor (BiSFET)\cite{banerjee_bilayer_2009}. The BiSFET is an ultra-low power device concept based on gated double-layer graphene surrounded by low-$\kappa$ dielectrics to produce an excitonic Bose Condensate. The BiSFET requires substantial reduction of the coupling between the two graphene layers, but coupling should not be entirely eliminated. One potential way to achieve this goal is to insert intercalants between graphene bilayers, or into graphite. This insertion of intercalant between successive layers of graphene forms what are known as graphene intercalation compounds (GICs)\cite{dresselhaus_intercalation_1981}. The choice of intercalant species and amount of intercalation provides a useful knob to control various properties like carrier concentration and interlayer electrical conductivity. Bilayer graphene (BLG) is made up of two separate layers of graphene stacked usually in AB and sometimes in AA configuration. Native BLG has a very strong inter-layer interaction and thus high c-axis conductivity making it unsuitable for the BiSFET. The presence of intercalant between the carbon layers makes their interaction indirect, thereby reducing its magnitude\cite{kulbachinskii_shubnikovhaas_1995}. In the case of BiSFET, achieving optimal coupling is critical for the formation of the condensate. GICs are also being considered for energy storage\cite{kumar_direct_2011}, nanoelectronics\cite{nathaniel_tunable_2012} and spintronics\cite{hiranandani_magnon_2012}. In order to design these novel devices it is critical to understand the structural and electronic properties of these intercalation compounds, specially inter-layer coupling. 

	In this paper we study intercalation of graphene with iodine monochloride (ICl) and iodine monobromide (IBr). ICl and IBr both form acceptor stage 1 intercalants with graphene\cite{dresselhaus_intercalation_1981} and are very promising for our devices. The I-Cl(Br) form covalent bonds through the p electrons\cite{pasternak_studies_1968} and graphene $\pi$ orbitals are modified by interaction with iodine $\pi$ orbitals. \cite{tiedtke_chemical_1990}. There is considerable charge transfer from graphene to the halogen molecules and also from iodine to chlorine(bromine) in ICl(IBr)-GIC. The reported c-axis conductivity is very low with an anisotropy ratio of $\rho_c/\rho_a = 10^3$\cite{ohta_c-axis_1986}. Sugihara\cite{sugihara_c-axis_1984} has proposed a theory of c-axis conduction in acceptor-type GICs via hopping and the low c-axis conductivity points to strong localization of $\pi$ electrons in the graphene layer. All of this makes ICl-GIC and IBr-GIC strong candidates for the formation of excitonic condensates in graphene double layers. We first address the question of structural configuration of these intercalation compounds. We observe that while the iodine sites proposed in earlier literature are correct, the chlorine and bromine sites are not. In fact there is a possibility that the intercalant is \textit{incommensurate} with graphene. Then, electronically, we observe hole-doping of graphene along with a small gap opening associated with breaking the sublattice symmetry within the individual graphene layers, as well as a smaller energy splitting associated with the graphene interlayer coupling. The interlayer coupling is mediated via the the overlap between C-p and the intercalant orbitals. The latter represents a drastic but not complete reduction of electronic coupling between the graphene layers. These results are promising towards the realization of a GIC based BiSFET. 
	
\section{Structural and Computational details}

	Iodine monochloride (ICl) and iodine monobromide (IBr) are molecular species that form acceptor stage 1 intercalants with graphene\cite{dresselhaus_intercalation_1981, pasternak_studies_1968}. While being similar to Br$_2$-GIC\cite{krone_intercalate_1989}, which has been well studied, the structural properties of these intercalation compounds are still subject to debate. In this paper we hope to illuminate that issue. In line with most of the literature on ICl-GIC, as well as our own experiments, we constructed models corresponding to vapor phase intercalation of graphene with ICl and IBr. Heerschap \textit{et al.} studied the microstructure associated with in-plane imperfections, noting that the C planes in graphite-ICl are AA stacked\cite{heerschap_electron_1964}. The in-plane structure for graphite intercalated with ICl was first reported by Turnbull \textit{et al.}\cite{turnbull_1966} in 1966. They studied diffraction patterns to conclude that ICl-to-graphite layer spacing is 3.65 \AA\ while the c axis parameter is 7.0 \AA. They also suggested a monoclinic lattice for the intercalant with lattice vectors a $\sim$ 58 \AA\ and b $\sim$ 28 \AA. In parallel they proposed a smaller \textit{pseudo-cell} structure that fit their diffraction results. This latter structure had parameters a = 4.92\AA, b = 19.2\AA\ and $\Gamma$ = 93.5\degree. While the positions of iodine were obtained by the Sayre-Cochran relation\cite{turnbull_1966}, the chlorine sites were obtained by trial and error and the assumption of centrosymmetry. This structure was however challenged by Moret \textit{et al.}\cite{moret_single-crystal_1983} in 1983, after they performed a high resolution photographic study of ICl-GIC suggesting that the structure was incommensurate. A different study done by Ghosh \textit{et al.}\cite{ghosh_superlattice_1984} in 1984 found that a much bigger unit cell fit the diffraction pattern better. This unit cell was nearly double the size of the Turnbull model along the b lattice vector. Wortmann, Krone and Kaindl\cite{wortmann_x-ray_1988, krone_intercalate_1989} followed up with a model which was a refinement of the Turnbull structure. They specified that there was parallel alignment of intercalated ICl molecules with respect to the graphite planes. Moreover the intra-molecular I-Cl bond length was 2.54 \AA, considerably larger than that in the gas or solid phase. Keeping the same iodine sites as Turnbull \textit{et al.} and using their new I-Cl bond length along with an assumption about inversion symmetry, they proposed a new pseudo-cell which was the basis of our initial structure in this study. 

\begin{figure}
\includegraphics[width=0.9\linewidth]{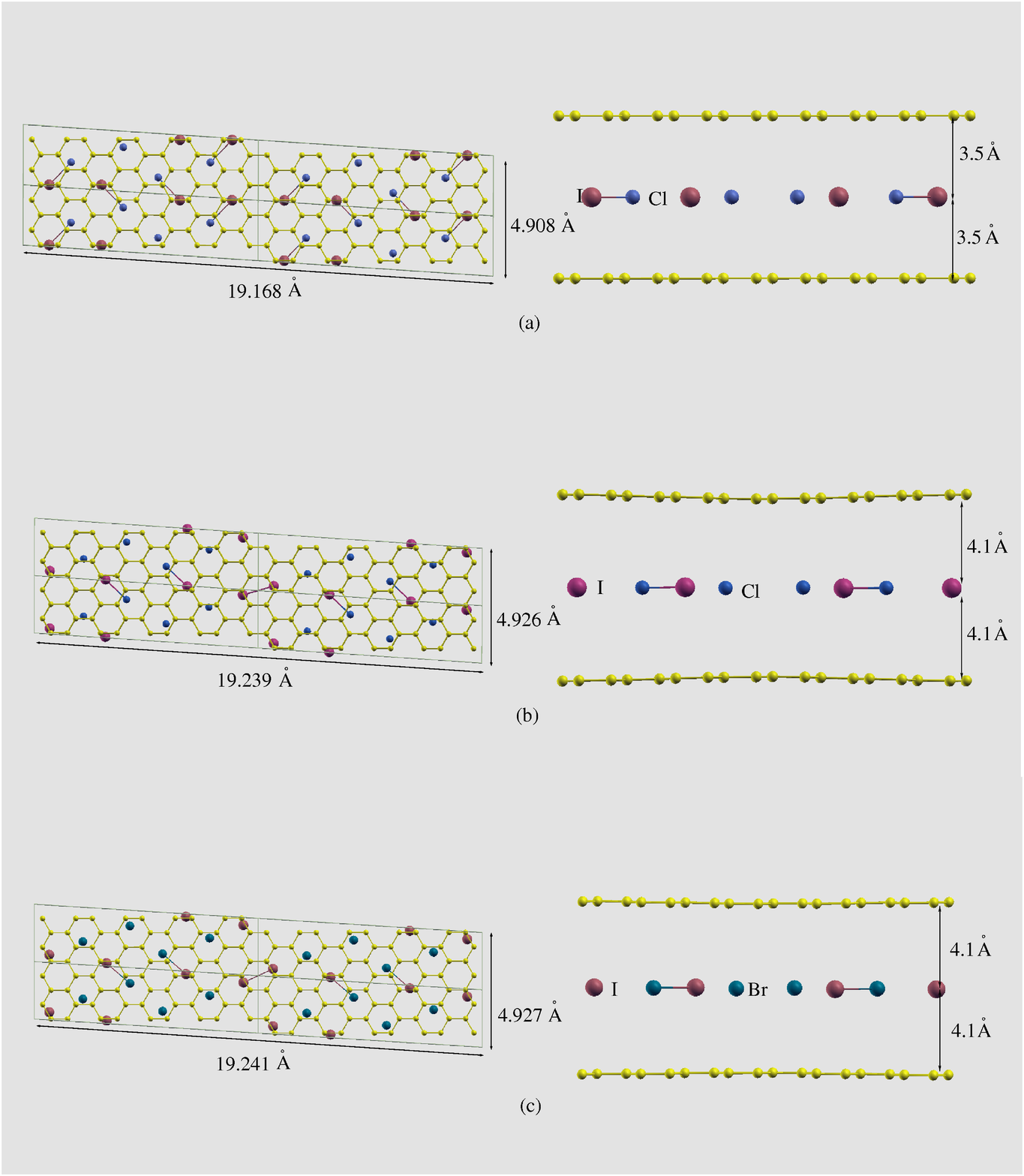}
  \caption{Model structures for ICl-GIC and IBr-GIC (a) Top and side view of ICl-GIC model before relaxation (b) Top and side view of ICl-GIC structure after relaxation (c) Top and side view of IBr-GIC structure after relaxation.}
  \label{fig1}
\end{figure}

	The density functional theory (DFT) calculations were done using the Vienna Ab-inito Simulation Package\cite{VASP1, VASP2, VASP3} with PAW pseudopotentials\cite{PAW} and a Generalised Gradient Approximation (GGA)\cite{GGA1} based exchange correlation correction. Our model constitued of 36 C atoms in one C layer with 4 ICl units per unit cell. Taking C-C bond length as 1.417 \AA\  we get the lattice parameters as $\vec{a}  = 19.1295 \vec{x} - 1.227 \vec{y}$, $\vec{b} = 4.908 \vec{y}$ and $\Gamma$ = 93.670\degree. We took the c lattice constant as 17 \AA\ with 10 \AA\ of vacuum between neighboring supercells. The distance between ICl and either C layer was 3.5 \AA\ with ICl bond length as 2.5544 \AA\ (\ref{fig1}). Starting with this structure we optimized the Kmesh to 2x8x2 and energy cut off to 450 eV while checking for convergence. Subsequently we relaxed the ions and optimized the in-plane lattice constants until the forces were smaller than 0.01 eV/\AA. 

	After relaxation, the structure obtained for ICl-GIC (\ref{fig1}) was slightly different from the starting structure. To begin with, there was a small volume expansion such that the lattice constants corresponding to the relaxed structure were 1.0036 times the initial lattice constants. The relaxed graphene layers displayed some buckling, as is evident from \ref{fig1}. The equilibrium interlayer distance between the two C layers is 8.2 \AA\ which is larger than the $\sim$7 \AA\ obtained experimentally. However, we can attribute this to the absence of Van der Waal's force in our simulations. The iodine sites in our relaxed structure however agree remarkably well with the Turnbull model. We also see evidence of inversion symmetry with 2 equivalent I and Cl sites in the unit cell which is in keeping with the spectra studied by Tiedke \textit{et al.}\cite{tiedtke_chemical_1990}. The I-Cl bond length is around 2.5 \AA\ and more. The Cl sites are quite different from those proposed by Turnbull \textit{et al.} or by Wortmann \textit{et al.}. This could point to the incommensurability of the ICl structure with the graphene plane.
	
	The reported structure and properties for IBr intercalation compounds are described as being very similar to those of the ICl intercalation compounds\cite{tiedtke_129i-mossbauer_1989, barati_electron-phonon_1999,chung_structural_1978}. Hence we started from the same initial structural model as used for ICl-GIC and replaced Cl with Br. The Kmesh was taken as 2x6x2 and energy cut off as 450 eV as they both sufficed for convergence of total energy. Relaxation of ions and optimization of lattice constants was performed using VASP with PAW pseudopotentials and GGA for exchange correlation approximation. The resulting structure is shown in \ref{fig1}. Again there was an expansion in the unit cell with the fractional change in lattice constants being 1.00382. This is larger than that seen for ICl-GIC. As a result the I-Br bond lengths are also larger than I-Cl, clustering around 2.7 \AA. The I and Br sites are very similar to the I and Cl sites, respectively, with the structure displaying inversion symmetry and 2 inequivalent site type for I and Br. We thus expect the electronic and transport properties of both these intercalated compounds to be qualitatively similar.

\section{Experimental results}

\begin{figure}
\includegraphics[width=0.9\linewidth]{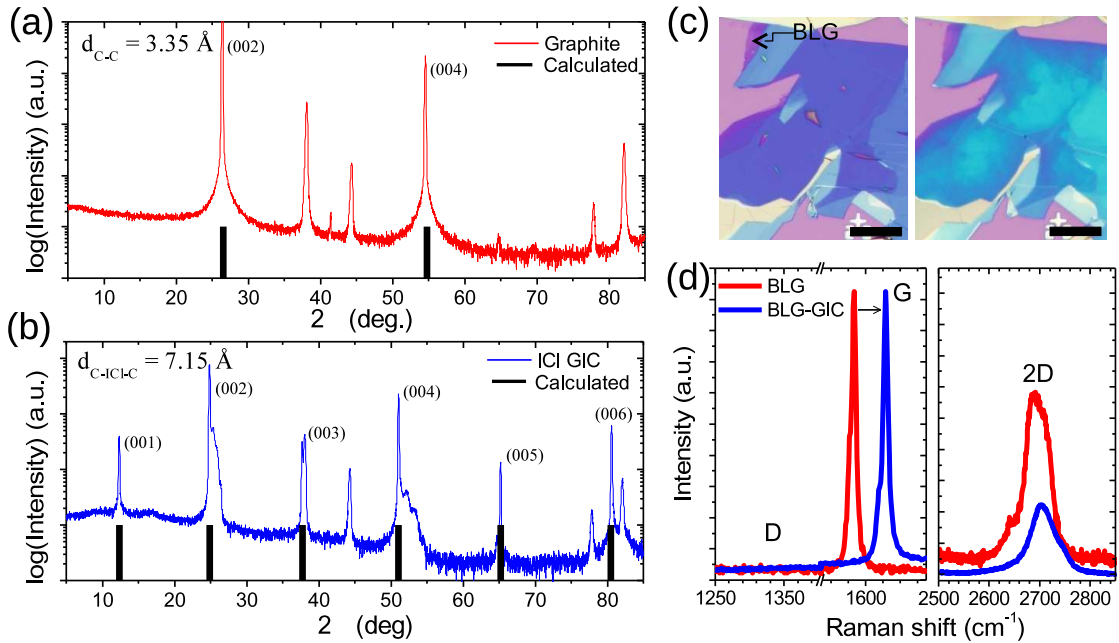}
  \caption{Experimental results. (a) $\theta-2\thetaθ$ XRD pattern of bulk-graphite, showing characteristic peaks that correspond to a c-axis spacing (d$_{C-C}$) of 3.35 \AA. The unlabeled peaks come from the Al substrate holder. (b) The XRD pattern of bulk-graphite stage 1 ICl-GIC shows additional peaks from the \textit{(001)}, \textit{(003)} and \textit{(005)} planes. The interlayer spacing is extracted as d$_{C-ICl-C}$ = 7.15 \AA. (c) Optical micrographs of an FLG flake before (left) and after (right) intercalation show changes in optical contrast. The scale bars are 50 $\mu$m. (d) Raman spectra of BLG and BLG-GIC show an upshift of the G-peak and a change in the 2D-peak shape towards a single Lorentzian after intercalation. The absence of D-peaks indicates defect free graphene.}
  \label{fig2}
\end{figure}

We used ICl to intercalate bulk-graphite and few layer graphene (FLG) into stage 1 ICl-GICs. X-ray diffraction (XRD) was used to determine the stage number of the bulk-graphite ICl-GIC. We extracted the c-axis spacing from the XRD peak positions, which increases from 3.35 \AA\ in graphite to 7.15 \AA\ in stage 1 ICl-GIC. We then intercalated FLG and bilayer graphene (BLG) flakes and used Raman spectroscopy to characterize the resulting BLG-GIC. We estimate a Fermi level shift of $\sim$ 0.7 eV for the BLG-GIC. This agrees very well with our DFT results. 

	Bulk-graphite ICl-GIC was prepared from natural graphite flakes (NGS Naturgraphit, $10-20$ mm flakes) and ICl (Sigma-Aldrich, reagent grade, 98\%) using a two-zone vapor-phase intercalation process\cite{dresselhaus_intercalation_1981}. A chunk of bulk-graphite ($\sim$ 0.1 g) and ICl ($\sim$ 0.5 g) were introduced into separate chambers of a two-chamber Schlenk tube which was then pumped for 10 minutes at a base pressure of $\sim 3 \times 10^{-3}$ Torr to remove any traces of water vapor from the ICl. The tube was then sealed and inserted in an oven maintained at 35\degree C for a period of 48 hours, to obtain stage 1 ICl-GIC\cite{ohta_c-axis_1986}. The glass tube was cooled in a refrigerator at 4\degree C for 30 minutes to condense the ICl vapors prior to removing the bulk-graphite ICl-GIC. A higher intercalation temperature and/or a shorter intercalation time were found to result in GICs of higher stage number and incomplete staging. XRD analysis was performed on the resulting ICl-GIC using high-speed Bragg-Brentano optics (CuK$_{\alpha}$) on a PANalytical X'Pert Pro MRD system. A symmetric $\theta-2\theta$ scan was performed to determine the stage number and the interplanar c-axis spacing using Bragg's law \ref{eq1}.

\begin{equation}
2d_{hkl} \sin\theta = n\lambda \label{eq1}
\end{equation}
							   
\ref{fig2}(a),(b) show the $\theta - 2\theta$ XRD scans for bulk-graphite and bulk-graphite ICl-GIC, respectively. A c-axis interlayer spacing (d$_{C-C}$) of 3.35 \AA\ is extracted for bulk-graphite by fitting the calculated peak positions from the \textit{(002)} and \textit{(004)} planes to their observed values. The d$_{C-C}$ value agrees well with prior reports from literature\cite{baskin_lattice_1955, murakami_m._debye-waller_1996}. The XRD scan of bulk-graphite ICl-GIC shows additional peaks between the main bulk-graphite peaks, corresponding to reflections from the \textit{(001)}, \textit{(003)} and \textit{(005)} planes, which is a clear signature of stage 1 GIC\cite{dresselhaus_intercalation_1981, murakami_m._debye-waller_1996}. The c-axis interlayer spacing (C/ICl/C) is extracted to be d$_{C-ICl-C}$ = 7.15 \AA, which has been reported earlier in literature\cite{murakami_m._debye-waller_1996}. The broadening of the intercalant peaks in \ref{fig2}(b) suggests that even though the GIC predominantly consists of a stage 1 superlattice structure, it also includes minority crystal phases of graphite and possible higher stage GIC subdomains, possibly due to incomplete intercalation\cite{shih_bi-_2011}. We believe that an optimization of the intercalation recipe would result in a more homogeneous stage 1 GIC formation.

	To further characterize stage 1 ICl-GICs, we intercalated FLG and BLG flakes exfoliated onto 285 nm SiO$_2$/Si substrates. The same recipe used for bulk-graphite was used to intercalate the FLG and BLG flakes. \ref{fig2}(c) shows optical micrographs of an FLG flake, before and after ICl intercalation. Thicker FLG flakes show a marked change in optical contrast, whereas contrast changes in thinner FLG and BLG flakes are negligible. BLG flakes, however, do get intercalated, as indicated by changes in Raman signatures after intercalation. \ref{fig2}(d) shows the Raman spectra of a BLG flake measured at 532 nm, before and after intercalation. There is a large upshift of the G-peak from $1586 cm^{-1}$ to $1623 cm^{-1} (+ 37 cm^{-1})$, which is indicative of intercalation\cite{song_raman_1976}. Such a large upshift of the G-peak cannot be due to charge-transfer doping from ICl adsorption alone, but must be due to intercalation of ICl into the BLG flake\cite{zhan_fecl3-based_2010}. Using work done by Das \textit{et al.}\cite{das_phonon_2009}, we estimate the Fermi level shift of the BLG-GIC to be $\sim$ 0.7 eV. This agrees very well with our theoretical predictions using DFT. The 2D-peak shape of BLG also changes significantly after ICl intercalation. Pristine BLG is AB stacked and exhibits a broad 2D-peak composed of four Lorentzian component peaks\cite{ferrari_raman_2006}. However, after intercalation, the 2D-peak shape changes significantly, towards a single Lorentzian. This signifies electronic decoupling of the graphene layers into an AA  stacking configuration\cite{zhao_intercalation_2011}. The faint shoulder on the 2D-peak at $\sim 2740 cm^{-1}$ could possibly be from un-intercalated islands in the BLG-GIC flake\cite{zhao_intercalation_2011}. In addition, there is no D-peak observed for the BLG-GIC, indicating defect-free graphene after intercalation. Raman spectra of FLG-GIC show a similar behavior, with a large upshift of the G-peak and a change in the 2D-peak shape towards a single Lorentzian.

\section{Electronic structure}
\begin{figure}
\includegraphics[width=0.9\linewidth]{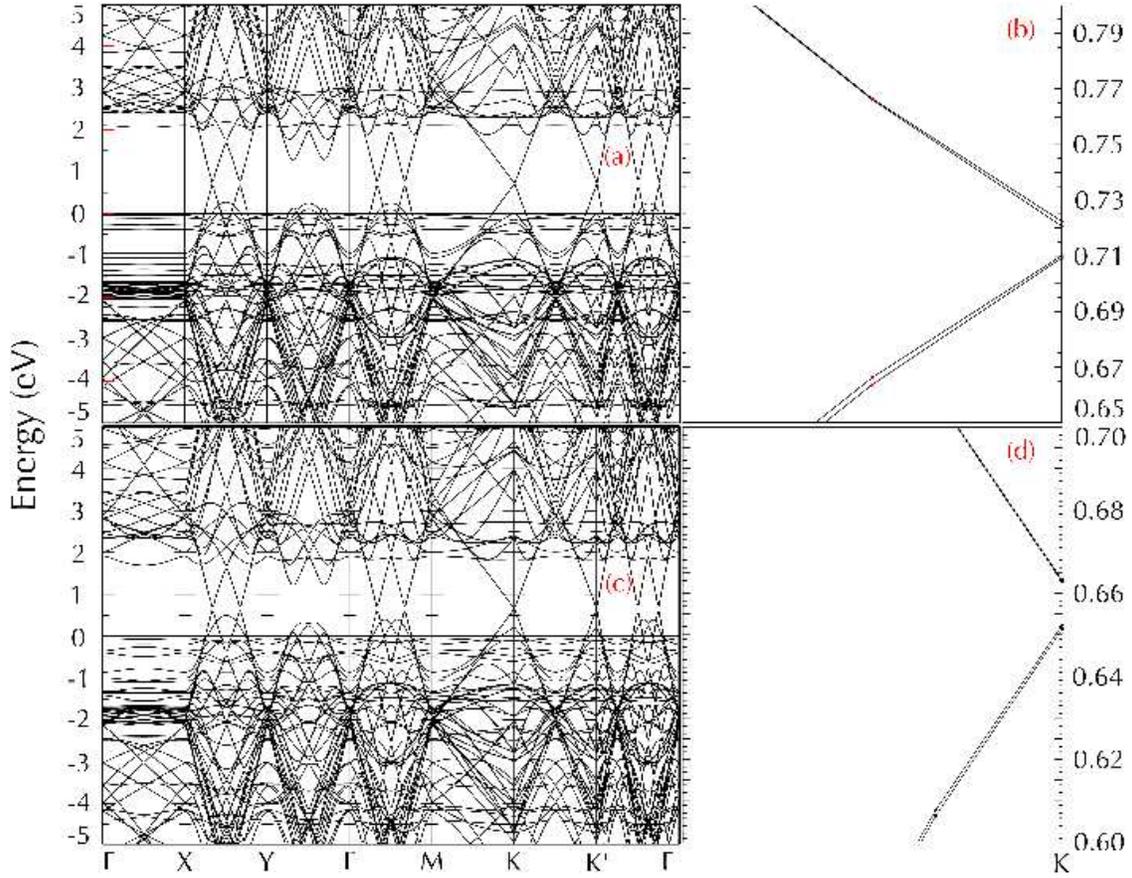}
  \caption{Electronic structure and gap opening at the K point for, (a) and (b), relaxed ICl-GIC, and, (c) and (d), relaxed IBr-GIC. The zero energy reference is the Fermi level. The M, K and K' points correspond to those of the Brillouin zone of the graphene layers.}
  \label{fig3}
\end{figure}

	We calculated the bandstructure of relaxed bilayer graphene intercalated with ICl and IBr. As seen in \ref{fig3}, graphene maintains its linear spectrum except very near the K point. However, it becomes hole doped due to the presence of the acceptor type intercalant. Moreover, for both structures there is a small gap between the (nominally) conduction and valence bands at the Dirac point, i.e. between the upper and lower cones, equal to 12 meV for ICl-GIC and 11 meV for IBr-GIC, which we label \textit{E$_{g1}$}. There is still smaller splitting throughout the conduction band and, separately, throughout the valence band, the value of which at the K point we label \textit{E$_{g2}$}. The value of \textit{E$_{g1}$} and, for each band, that of \textit{E$_{g2}$} are provided in \ref{table1}. 
	
\begin{figure}
\includegraphics[width=0.9\linewidth]{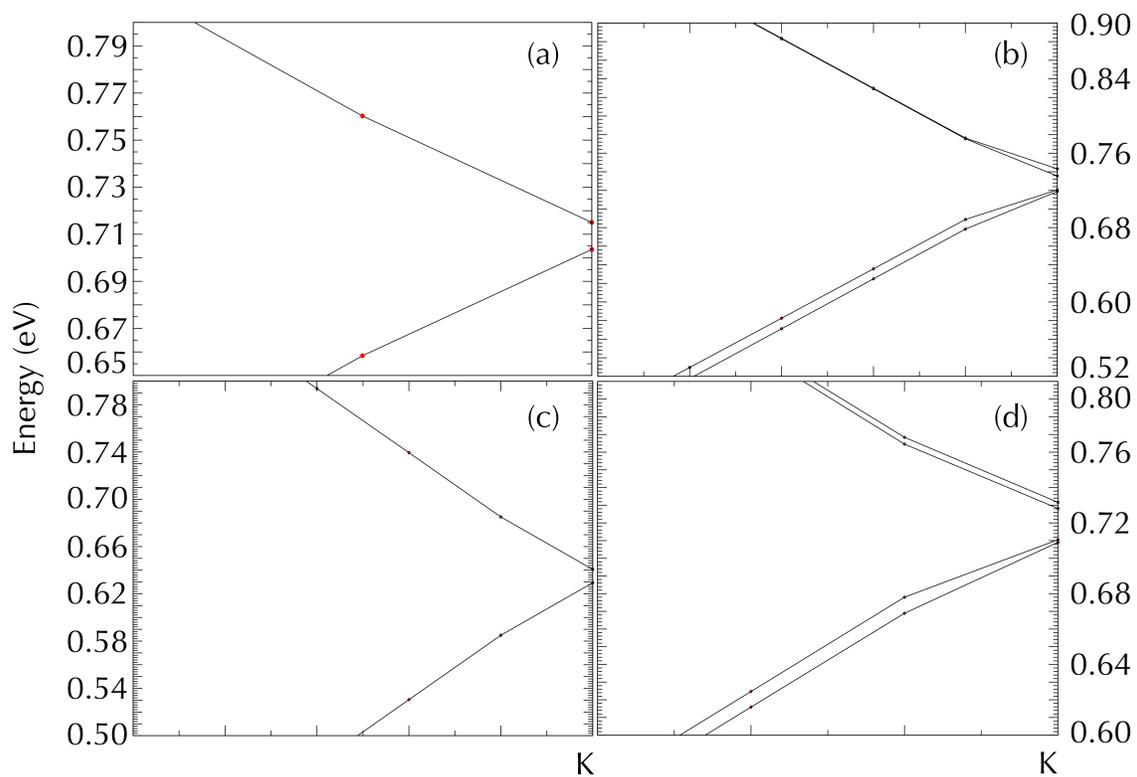}
  \caption{Gap opening at the K point for (a) relaxed C-ICl (b) ICl-GIC with the experimental C-C distance (d$_{C-C}$) of 7.15\AA\ (c) relaxed C-IBr (d) IBr-GIC with the experimental C-C distance of 7.15\AA. The zero energy reference is the Fermi level. The K point corresponds to that of the Brillouin zone of the graphene layer(s).}
  \label{fig4}
\end{figure}

There are two sources of degeneracy breaking here, that of coupling between the graphene layers and that of symmetry breaking between the sublattices of the individual layers due to coupling to the intercalant. To help isolate the effects of each, we performed band structure calculations in structures with one of the two graphene layers removed, i.e. for just C-ICl and and C-IBr as shown in \ref{fig4}(a) and \ref{fig4}(c), respectively. The results, with only the valence-to-conduction band splitting, show coupling to the intercalant to be the source of \textit{E$_{g1}$} by its presence, and weakened interlayer coupling between the graphene layers through the intercalant to be the source of the \textit{E$_{g2}$} splitting by its absence. The coupling between the graphene layers is of particular importance for the proposed BiSFET. It is well known that the two C layers in AB stacked bilayer graphene are highly coupled, which leads to their parabolic band structures near the K points and diminishes the symmetry between valence and conduction bands\cite{xu_infrared_2010} close to the Fermi level. In AA stacked bilayer graphene this C-C coupling leads to the two Dirac points split above and below the Fermi level. The presence of intercalant molecules makes the C-C interaction indirect, i.e. mediated through overlap with the intercalant molecules\cite{kulbachinskii_shubnikovhaas_1995}. This significantly reduces the magnitude of this intercation as compared to the bilayer case. This reduced coupling, mediated through hopping between C-p and intercalant orbitals, in principle determines the critical (maximum) interlayer condensate transport in the BiSFET. We also note that band structure obtained for methane-intercalated bilayer graphene\cite{hargrove_band_2012} for both AA and AB stacking turn out to be very similar. Our results for AA stacking can thus be qualitatively extended to intercalated AB stacked graphene, probably due to indirect interaction between the C layers. However, as previously noted, the interlayer distances we obtained during DFT-based relaxation of the intercalated structure were obtained without Van der Waal's force. Because the electronic coupling between the layers is strongly dependent on the interlayer distance, we also conducted a set of calculations with the intercalated carbon layers "squeezed" down from the "relaxed" 8.2 \AA\ to the experimental value of 7.15 \AA\, and with half the latter between the C and the intercalant layers when considering just one graphene layer. These latter results are summarized in \ref{table1}, and the band structures for the intercalated structures are shown in \ref{fig4}(b) and (c). We found, as qualitatively expected, that forcing the layers closer both increased the interactions leading to a higher \textit{$E_{g1}$} and \textit{$E_{g2}$} and somewhat increased the doping of the C layers.

\begin{table}
\caption{Gap openings between intercalated graphene energy bands due to interaction of the graphene layers with the intercalant and with each other. \textit{E$_{g1}$} is the separation between the (nominally) conduction and valence bands; the two values of \textit{E$_{g2}$} are the splitting of the conduction band and of the valence band, respectively, at the K point. All gaps are in meV. "Relaxed" structures are obtained upon DFT relaxation absent Van der Waal's force; "squeezed" structures have the layer separations based on the experimental intercalated graphene layer distance artificially of 7.15\AA.}
  \label{table1}
  \begin{tabular}{|c|c|c|c|c|c|c|}
\hline
System	&\multicolumn{2}{|c|}{ICl-GIC}	&\multicolumn{2}{|c|}{IBr-GIC}	&C-ICl	&C-IBr\\
\hline
	&\textit{E$_{g1}$}	&\textit{E$_{g2}$}			&\textit{E$_{g1}$}	&\textit{E$_{g2}$}			&	&	\\
\hline
Relax	&12	&1,2			&11	&1,1			&12	&12	\\
\hline
Squeezed &16	&2,8			&20	&2,4			&21	&17	\\
\hline
  \end{tabular}
\end{table}

\section{Conclusion}
In this paper we have presented first-principles density functional theory based calculations of bilayer graphene intercalated with iodine monochloride and iodine monobromide forming stage 1 acceptor-type intercalation materials. These systems are highly promising for the proposed untra-low power BiSFET\cite{banerjee_bilayer_2009}, as well as perhaps other novel devices\cite{kumar_direct_2011, nathaniel_tunable_2012, hiranandani_magnon_2012}. We have studied the structural configurations of these materials which have been subject to debate. Moreover, we have explored the electronic properties of ICl-GIC and IBr-GIC, specially focussing on the inter-layer interactions which are highly critical for novel device functionality. In addition, we have presented experimental results on intercalation of bilayer graphene (BLG) and few layer graphene (FLG) with ICl.

\begin{acknowledgement}
We thank NRI-SWAN for financial support and the Texas Advanced Computing Center (TACC) for access to high performance computing resources.
\end{acknowledgement}

\bibliography{ICl_IBr_GIC}

\end{document}